\documentclass[11pt]{article}
\usepackage{amsmath}
\usepackage{latexsym}

\allowdisplaybreaks

\newcommand{\be}{\begin{equation}}
\newcommand{\ee}{\end{equation}}
\newcommand{\bea}{\begin{eqnarray}}
\newcommand{\eea}{\end{eqnarray}}
\newcommand{\nn}{\nonumber} 
\newcommand{\p}{\partial} 

\newcommand{\cV}{\mathcal{V}}

\newcommand{\cN}{\mathcal{N}}

\newcommand{\tr}{\mathrm{tr}}

\newcommand{\tp}{\chi}

\newcommand{\I}{\mathrm{i}}

\begin{document}
\begin{titlepage}
\begin{flushright} \small
 ITP-UU-07/05 \\ SPIN-07/05 \\ hep-th/0701223
\end{flushright}
\bigskip
\renewcommand{\thefootnote}{\fnsymbol{footnote}}
\begin{center}
 {\LARGE\bfseries Tensor supermultiplets and \\[1.1ex] toric
  quaternion-K\"ahler geometry\footnote[2]{Talk given by F.S.\ at the RTN
    ForcesUniverse Network Workshop, Napoli, October 9th - 13th,
    2006.}}
\\[10mm]
\textbf{Bernard de Wit and Frank Saueressig}\\[5mm]
{\em Institute for Theoretical Physics} and {\em Spinoza
  Institute,\\ Utrecht University, Utrecht, The Netherlands}\\[3mm] 
{\tt  B.deWit@phys.uu.nl}\;,\; 
{\tt F.S.Saueressig@phys.uu.nl}
\end{center}
\renewcommand{\thefootnote}{\arabic{footnote}}
\setcounter{footnote}{0}
\vspace{5mm}

\bigskip

\centerline{\bfseries Abstract} 
\medskip
\noindent
We review the relation between $4n$-dimensional quaternion-K\"ahler
  metrics with $n+1$ abelian isometries and superconformal theories of
  $n+1$ tensor supermultiplets. As an application we construct the
  class of eight-dimensional quaternion-K\"ahler metrics with three
  abelian isometries in terms of a single function obeying a set of
  linear second-order partial differential equations. \medskip
\end{titlepage}
\section{Introduction}
\label{sec:intro} 
A fascinating feature of string theory is the rich interplay between
its physical properties and mathematics.  An example of this is
provided by supersymmetric effective actions, where the scalar fields
coordinatize a target space manifold whose geometry is of a restricted
type.  Based on this close link, it is not surprising that progress in
understanding the low-energy limit of string theory often comes hand
in hand with a better understanding of the corresponding geometries.

This paper focuses on the geometries arising in the hypermultiplet
sector of $N=2$ supergravity theories. Since a hypermultiplet contains
four real scalars, the corresponding manifolds have dimension $4n$,
where $n$ is the number of hypermultiplets.  Supersymmetry restricts
the underlying metric to be quaternion-K\"ahler \cite{BW}. In case of
a single hypermultiplet, this is equivalent to the metric being
Einstein with self-dual Weyl curvature. These spaces have been studied
extensively by both mathematicians and physicists (see {\it e.g.}
\cite{CP,Tod,Prz}) and it turned out that the most general
four-dimensional quaternion-K\"ahler metric is determined by a single
function obeying a non-linear partial differential equation
\cite{Prz}. When the metric is toric, {\it i.e.}, when it has two
commuting isometries, this equation can be linearized and reduces to a
Laplace-like equation on the upper half-plane \cite{CP}. 
For $n > 1$ the quaternion-K\"ahler manifolds are characterized by
their holonomy group $\mathrm{Sp}(1)\times\mathrm{Sp}(n)$.  This makes
the study of their metrics very involved since a generic
quaternion-K\"ahler manifold does, for instance, not admit a K\"ahler
potential.

{From} $N=2$ supergravity theories in four space-time dimensions
\cite{DVV} it is clear that there exists a relation between
quaternion-K\"ahler manifolds and hyperk\"ahler cones. The latter
provide the hypermultiplet target spaces of field theories that
are invariant under the rigid $N=2$ superconformal symmetry
\cite{superconformalhypers}.  Such a cone has a homothetic Killing
vector and three complex structures which rotate isometrically under
the group $\mathrm{Sp}(1)$ \cite{Swann,Galicki}.  It is a cone over a
$(4n+3)$-dimensional 3-Sasakian manifold, which in turn is an
$\mathrm{Sp}(1)$ fibration of a $4n$-dimensional quaternion-K\"ahler
space. The quaternion-K\"ahler manifold is the $N=2$ superconformal
quotient of the hyperk\"ahler cone and there is a one-to-one relation
between quaternion-K\"ahler spaces and hyperk\"ahler cones.  This
relation was worked out explicitly in \cite{deWit:2001dj}.

Quaternion-K\"ahler geometries simplify further when the corresponding
hyperk\"ahler cone possesses $n+ 1$ abelian symmetries. In this case
one can use the (Hodge) duality between scalars and second rank tensor
gauge fields to dualize some of the scalars into tensors and work with
tensor supermultiplets instead of hypermultiplets. The conformal
supergravity theories including an arbitrary number of tensor
supermultiplets and their corresponding superconformal quotient have
recently been worked out in \cite{deWit:2006gn}. Subsequently this
formulation has led to major progress in understanding perturbative
and non-perturbative properties of four-dimensional effective string
actions \cite{RSV,RRSTV}.

Here we focus on the mathematical implications of the superconformal
quotient for superconformal tensor multiplets coupled to supergravity.
In \cite{deWit:2006gn} this technique has already been used to give an
elegant derivation of the four-dimensional quaternion-K\"ahler metrics
with two commuting isometries, obtained originally in \cite{CP}.
After reviewing the general framework, we will construct the class of
eight-dimensional quaternion-K\"ahler metrics with three abelian
isometries. These metrics are described by a single function which
obeys a simple set of linear partial differential equations. Further
details and proofs will be presented in a forthcoming paper.

\section{Superconformal tensor multiplet Lagrangians}
\label{sec:tensor-lagrangians}
The bosonic fields of an $N=2$ tensor supermultiplet consist of three
scalar fields, an antisymmetric tensor gauge field $E_{\mu \nu}$, and
a complex auxiliary field (the latter will be ignored in the following).
The scalar fields transform in the vector representation of the
$\mathrm{SU}(2)$ R-symmetry group, and can be described conveniently
in terms of symmetric tensors $L^{ij}$, satisfying $L_{ij}=
(L^{ij})^\ast = \varepsilon_{ik} \, \varepsilon_{jl} \, L^{kl}$. In
\cite{deWit:2006gn} we proved that superconformal Lagrangians of
$n+1$ tensor multiplets are encoded in a ``potential'' $\chi(L)$, which
depends on the fields $L_{ij}{}^I$ labelled by indices
$I=1,2,\ldots,n+1$. Conformal invariance requires this potential to be
$\mathrm{SU}(2)$ invariant and homogeneous of degree $+1$, which
implies
\begin{equation}
  \label{eq:chi-su2-D}
  \frac{\partial\chi(L)}{\partial L^{ij I}} \,L^{jk I} = \tfrac12
  \,\delta_i{}^k \, \chi(L) \,. 
\end{equation}
Supersymmetry requires the second derivative of $\chi(L)$ to satisfy
the condition,
\begin{equation}
  \label{eq:chi-metric} 
    \varepsilon_{kl}\;
  \frac{\partial^2\chi(L)}
  {\partial  L_{ik}{}^I\,\partial  L_{jl}{}^J}= 2\, F_{IJ}(L) 
  \;\varepsilon^{ij} \,,
\end{equation}
which defines a function $F_{IJ}(L)$, symmetric in $I$ and $J$. As it
turns out $F_{IJ}(L)$ plays the role of the target space metric of the
$3(n+1)$ tensor multiplet scalars. One can prove that (\ref{eq:chi-su2-D})
and (\ref{eq:chi-metric}) imply two more equations, 
\begin{equation}
  \label{eq:Laplace-F} 
    \varepsilon_{ij} \,\varepsilon_{kl}\;
  \frac{\partial^2 F_{IJ}(L)}
  {\partial  L_{ik}{}^K\,\partial  L_{jl}{}^L} = 0\,, \qquad
  \chi(L) = 2 F_{IJ}(L) \,L_{ij}{}^I L^{ijJ}\,,
\end{equation}
the first of which appeared originally in
\cite{Lindstrom:1983rt,Hitchin:1986ea}. As a by-product of the above
equations one deduces that the $k$-th derivative of $\chi(L)$ is
symmetric in both the multiplet indices $I_1,I_2,\ldots, I_k$ and in
the $\mathrm{SU}(2)$ indices $i_1,i_2,\ldots,i_{2k}$, and can thus be
denoted unambiguously by $\chi_{I_1\cdots I_k}{}^{i_1\cdots i_{2k}}$.

Because of $\mathrm{SU}(2)$ invariance, the potential $\chi(L)$
depends on only $3n$ independent variables and is subject to
$\tfrac32 (n-1)(n+2)$ second-order differential equations (for
$n\geq2$) corresponding to (\ref{eq:chi-metric}). For $n=1$,
$\chi(L)$ depends on 3 variables and is subject to a single
differential equation, while, for $n=0$, $\chi(L)$ is proportional to
$\sqrt{L_{ij}L^{ij}}$.

The general Lagrangian coupling an arbitrary number of tensor
multiplets to conformal supergravity has recently been worked out in
\cite{deWit:2006gn}. Here we display its bosonic part without the
auxiliary tensor multiplet fields, 
\begin{eqnarray}
  \label{eq:SCL}
  e^{-1} \mathcal{L} & = & \chi \left[ \tfrac{1}{6} R + \tfrac{1}{2} D
  \right] + F_{IJ} 
  \left[ - \tfrac{1}{2} \mathcal{D}_\mu L_{ij}{}^I \mathcal{D}^\mu
  L^{ijJ} + E_\mu{}^I 
    E^{\mu J} - E^{\mu I} \cV_\mu{}^i{}_j L_{ik}{}^J
    \varepsilon^{jk}\right] \nonumber \\ 
  && + \tfrac{1}{2}\mathrm{i} \, e^{-1} \varepsilon^{\mu \nu \rho \sigma} \,
  F_{IJK}{}^{ij} \, \p_\rho L_{ik}{}^J \, \p_\sigma L_{jl}{}^K \,
  \varepsilon^{kl} \, E_{\mu \nu}{}^I \,,
\end{eqnarray} 
where the $E^{\mu I}$ denote the field strengths associated with the
tensor gauge fields. We also note the presence of some of the
conformal supergravity fields contained in the Weyl multiplet, such as
the vierbein field $e_{\mu}{}^a$, the $\mathrm{SU}(2)$ gauge fields
$\mathcal{V}_\mu{}^i{}_j$ and an auxiliary scalar field $D$.
Furthermore, $\mathcal{D}_\mu L_{ij}{}^I$ denotes a derivative
covariant with respect to local dilatations and local
$\mathrm{SU}(2)$, and $R$ is the Ricci scalar associated with a
spin-connection field $\omega_\mu{}^{ab}$.

An intricate term of the Lagrangian \eqref{eq:SCL} is the last one,
which is not manifestly invariant under tensor gauge transformations.
Tensor gauge invariance requires that $E_{\mu\nu}{}^I$ couples to a
tensor which is the pull back of a closed two-form. In addition,
supersymmetry requires the supersymmetry variation of this form to be
exact. The relevant two-form, $F_I \equiv F_{IJK}{}^{ij}(L) \,
\mathrm{d} L_{ik}{}^J \wedge \mathrm{d} L_{jl}{}^K \,
\varepsilon^{kl}$, is indeed closed, as well as exact under
arbitrary variations $L^I\to L^I+\delta L^I$.  This implies that it
can locally be expressed in terms of a one-form, $F_I = \mathrm{d}
A_I$. This one-form is, however, no longer invariant under
$\mathrm{SU}(2)$.

To deal with this problem one decomposes the scalar fields $L_{ij}$
according to a $\mathrm{U}(1)$ subgroup of $\mathrm{SU}(2)$ into a
real and a complex field, $x$ and $v$, respectively
\cite{Lindstrom:1983rt,Hitchin:1986ea} (see also,
\cite{deWit:2001dj,deWit:2006gn}), and we define $L_{12}{}^I =-
\tfrac{1}{2} \mathrm{i} x^I$, $L_{22}{}^I=v^I$. In these variables
(\ref{eq:chi-metric}) takes the form $\chi_{x^Ix^J} + \chi_{v^I\bar
  v^J} = 2 F_{IJ}$. Furthermore, one can show that $F_{IJ}$ can be
written as the second derivative of another, homogeneous, function
$\mathcal{L}(x,v,\bar v)$, {\it i.e.}, $F_{IJ} =
\mathcal{L}_{x^Ix^J} = -\mathcal{L}_{v^I\bar v^J}$. The function
$\mathcal{L}$ is only invariant under a $\mathrm{U}(1)$ subgroup of
$\mathrm{SU}(2)$, and its multiple derivatives are again symmetric in
the multiplet indices $I,J,K,\ldots$.\footnote{
  The function $\mathcal{L}(x,v,\bar v)$ can elegantly be expressed
  as a contour integral over an auxiliary complex variable $\zeta$, 
  \begin{equation} 
    \label{eq:contour} 
    \mathcal{L}(x, v, \bar v) = \mathrm{Im} \, \oint \frac{\mathrm{d}
      \zeta}{2\pi \I \zeta} \, H(\eta^I) \, , 
  \end{equation}
  where $\eta^I = v^I / \zeta + x^I - \bar v^I \zeta$ and $H(\eta^I)$ is an
  arbitrary function homogeneous of degree one.
} 
 The potential $\chi$ can be expressed in terms of $\mathcal{L}$ by  
\begin{equation}
  \label{eq:chiF}
  \chi = \mathcal{L} - x^I \mathcal{L}_{x^I} \, .
\end{equation} 
Another relation is provided by the second equation of
(\ref{eq:Laplace-F}). 

In terms of $\mathcal{L}$ we can write down an explicit representation
for the one-form $A_I$,
\begin{equation}
  \label{eq:A-potential}
  A_{I} = \mathrm{i} \left( \mathcal{L}_{x^I \bar v^J} \mathrm{d}\bar v^J
  - \mathcal{L}_{x^I v^J} \mathrm{d}v^J \right) \, .   
\end{equation}
The coefficients $\mathcal{L}_{x^Iv^J}$ can be found from the
potential $\chi(L)$ by integration, as their first derivative with
respect to $x^K$ is symmetric in $I,J,K$ and equal to
$\mathcal{L}_{x^Ix^Jv^K} = \tfrac1{2} (\chi_{x^Ix^Jv^K}+ \chi_{\bar
  v^I v^J v^K})$. 

We stress that the Lagrangian (\ref{eq:SCL}) can be expressed in
terms of both $\mathcal{L}$ and $\chi$. Which formulation is more
convenient depends on the problem to be addressed. 

\section{The superconformal quotient}
\label{sec:quotient}
The superconformal Lagrangians considered in the previous section are
invariant under local superconformal symmetries. These theories are
gauge equivalent to matter-coupled supergravity of the standard
Poincar\'e type.  The conversion, which is known as the superconformal
quotient, can be effected either by imposing gauge conditions or by
employing gauge invariant variables so that the gauge degrees of
freedom will eventually decouple from the Lagrangian. Part of the
superconformal symmetries reflect themselves in the Lagrangian
(\ref{eq:SCL}), namely the dilatations and local $\mathrm{SU}(2)$
transformations. They imply that the target space for the scalar
fields is a cone. After performing the quotient and upon dualizing the
tensor fields to scalars, one ends up with a $4n$-dimensional
quaternion-K\"ahler space.\footnote{
  Without the fields of the Weyl multiplet, the superconformal
  symmetries are only realized in a rigid way. The resulting target
  spaces, after a tensor-scalar duality transformation, are
  $4(n+1)$-dimensional hyperk\"ahler cones
  \cite{deWit:2001dj,superconformalhypers}.
} 
For the Lagrangian \eqref{eq:SCL} the superconformal quotient has been
carried out in \cite{deWit:2006gn} and we will briefly summarize the
result. In the next section we will demonstrate the results for $n=2$.
The case $n=1$ was already treated in \cite{deWit:2006gn}. 

To carry out the superconformal quotient in the tensor sector, one
first converts to scale invariant variables by rescaling the
$L_{ij}{}^I$ by the inverse of $\chi$ and switches from the symmetric
tensor to a vector notation according to,
\begin{equation}
  \label{eq:tensor>vector}
  L_{ij}{}^I = - \mathrm{i} \chi\, \vec L{}^I \cdot (\vec\sigma)_i{}^k
  \,\varepsilon_{jk} \,.   
\end{equation}
The scale invariant fields are then subject to a
non-trivial constraint, $F_{IJ}(\vec L) \,\vec L^I\cdot \vec L^J =
\tfrac14$, where $F_{IJ}(\vec L) = \chi\,F_{IJ}(L)$. Here we
make use of the homogeneity of $F_{IJ}$. 

Subsequently, the $\mathrm{SU}(2)$ gauge fields
$\mathcal{V}_\mu{}^i{}_j$ are eliminated through their equation of
motion and the Lagrangian \eqref{eq:SCL} takes the form, 
\begin{eqnarray}  
  \label{eq:tensor-boson-noV}
  e^{-1} \mathcal{L} &=& 
  \chi \Big[ \tfrac{1}{6} R  
     +  \tfrac1{2} D  -\tfrac1{4} (\partial_\mu
    \ln \chi)^2 \Big] 
    \nonumber
\\[1ex]
&&{}
    - \chi\;
    \Big[\mathcal{G}^{(1)}_{IJ} \,\partial_\mu\vec L^I \,
    \partial^\mu\vec L^{J} + 
    \mathcal{G}^{(2)}_{IJ,KL} \, (\vec L^I\cdot\partial_\mu \vec L^J) \, 
    (\vec L^K\cdot\partial_\mu \vec L^L) \Big] 
    \nonumber
\\[1ex]
    &&{}
    + \chi^{-1}\,\Big[ \mathcal{H}^{(1)}_{IJ}\,
    E_{\mu}{}^I \, E^{\mu J} \Big]  
    + E^{\mu I} \, \Big[ \mathcal{H}^{(2)}_{IJ}\, \vec L^J\cdot(\vec
    L^K\times\partial_\mu\vec L^L) \, F_{KL} + A_{\mu I} \Big]\,. 
\end{eqnarray}
Note that in this notation, the scalar fields obviously parameterize a
cone with radial variable $\chi$. Eliminating the
$\mathrm{SU}(2)$ gauge fields does not affect the local
$\mathrm{SU}(2)$ invariance, so that the fields $\vec L^I$ comprise
$3n-1$ degrees of freedom (for $n\geq2$). Note that we have dropped a
total derivative as compared to (\ref{eq:SCL}), by making use of
(\ref{eq:A-potential}). As a result the tensor gauge fields appear
exclusively through their corresponding field strengths. The local
$\mathrm{SU}(2)$ invariance of the last term is realized because
$A_{\mu I}$ transforms nontrivially under $\mathrm{SU}(2)$.

To give the explicit expressions for the functions appearing above it
is useful to introduce the expression $\mathcal{N}_{I}{}^J =4\, F_{IK}
\vec L^K \cdot \vec L^J$, so that $\mathrm{tr}(\mathcal{N}) = 1$. When
lowering the upper index of the $k$-th power of this matrix by
contraction with $F_{IJ}$, one obtains a symmetric matrix
$\mathcal{N}^k_{IJ}$. Using this notation we obtain,
\begin{eqnarray}
  \label{eq:FHG}
    \mathcal{H}^{(1)}_{IJ} &= & F_{IJ} + F_{IK} \, L_r{}^{K} \,
    (\mathbf{M}^{-1})^{rs} \, L_s{}^{L} \, F_{LJ}\,,  \\[1mm] 
    \mathcal{H}^{(2)}_{IJ} &=& \frac{1}{16\, \det(\mathbf{M})} 
    \left[ \big(1  - \tr(\mathcal{N}^2)\big) \, F_{IJ} + 2
    \,\mathcal{N}_{IJ}^2 \right] \,,  \nn  \\    
    \mathcal{G}^{(1)}_{IJ} &= & 
    F_{IJ} - \frac{1}{128\, \det(\mathbf{M})} \left[ \big(1 +
    \tr(\mathcal{N}^2)\big) 
    \, \mathcal{N}_{IJ} - 2 \, \mathcal{N}_{IJ}^3 \right] \,, \nn \\ 
    \mathcal{G}^{(2)}_{IJ,KL} &= & 
    \frac1{16\, \det(\mathbf{M})} \,
    \left[ \mathcal{N}_{IK} \mathcal{N}_{JL} + \tfrac{1}{2} \big(1 +
    \tr(\mathcal{N}^2)\big) 
    F_{IL} F_{JK} - \big( \mathcal{N}^2_{IL} \, F_{JK} +
    \mathcal{N}^2_{JK} \, F_{IL} \big) \right]\,. \nn 
\end{eqnarray}
The $3\times  3$ matrix $\mathbf{M}$ arises naturally when eliminating
the $\mathrm{SU}(2)$ gauge fields and is given by 
\begin{equation}
  \label{eq:Mmat}
  \big[ \mathbf{M} \big]^{rs} = \left( \tfrac{1}{4} \, \delta^{rs} - L^{Ir} \,
  F_{IJ} \, L^{Js} \right) \,.
\end{equation}
Its determinant and inverse are equal to 
\begin{eqnarray} 
  \label{eq:det-M-inv} 
  \det(\mathbf{M}) &=& \tfrac{1}{192} \, \left[ 1 - \tr(\mathcal{N}^3 ) \right]
  \, , 
  \nonumber\\ 
  \big[ \mathbf{M}^{-1} \big]_{rs} &=& \frac{1}{32\,\det(\mathbf{M})} 
  \left[ \big(1 - \tr(\mathcal{N}^2) \big) \delta_{rs} + 8\, L_r{}^I
  \, \mathcal{N}_{IJ}\, L_s{}^J \right] \, . 
\end{eqnarray} 

\section{The case of three tensor multiplets}
\label{sec:3-multiplets} 
We now use the formalism of the previous section to construct
eight-dimensional quaternion-K\"ahler metrics with three abelian
isometries.  These isometries originate from the tensor fields, which
are dualized to scalars. This implies that we should start from
(\ref{eq:tensor-boson-noV}) with $n+1=3$ tensor
supermultiplets.\footnote{
  For compact spaces the maximal number of commuting isometries of a
  $4n$-dimensional quaternion-K\"ahler space equals $n+1$. In
  principle, the superconformal quotient can be applied to compact
  spaces, although only the non-compact spaces are physically relevant
  in supergravity. It is possible that non-compact spaces have more
  than $n+1$ abelian isometries, and these are not obvious in the
  context of our construction. Nevertheless we expect that the results
  of this paper cover the full class of quaternion-K\"ahler spaces
  with at least $n+1$ abelian isometries. This has been shown to be
  the case for $n=1$ \cite{deWit:2006gn,CP}.
} 

The natural starting point of the construction is the most general
potential $\chi$ for three tensor multiplets, $ L_{ij}{}^I$, where the
indices $I,J=1,2,3$ enumerate the tensor multiplets. The potential
must be invariant under $\mathrm{SU}(2)$ rotations. In order to make
this invariance manifest, we introduce the $\mathrm{SU}(2)$ invariant
fields,
\begin{equation}
  \label{eq:phi-var}
  \phi^{IJ} =   L_{ij}{}^I \, L^{ij J} \,,
\end{equation}
so that we are dealing with symmetric $3\times3$ tensors $\phi^{IJ}$.
In terms of these fields the constraints (\ref{eq:chi-su2-D}) and
(\ref{eq:chi-metric}) reduce to
\begin{equation}
  \label{eq:phi-const} 
  \phi^{IJ} \frac{\partial \chi(\phi)}{\partial\phi^{IJ}} = \tfrac12 
  \,\chi(\phi) \,, \qquad 
  \frac{\partial \chi(\phi)}{\partial\phi^{I[K}\, \partial \phi^{L]J}}
  = 0   \, .  
\end{equation} 
Furthermore we derive,
\begin{equation}
  \label{eq:FIJ}
  F_{IJ}(\phi) = \frac{\partial \chi(\phi)}{\partial \phi^{IJ}}\,.
\end{equation} 

Using these results we can evaluate the general formulae
\eqref{eq:tensor-boson-noV}, \eqref{eq:FHG} and \eqref{eq:Mmat} for
the case of three tensor multiplets. As before we introduce scale
invariant variables, $\hat \phi{}^{IJ} = \chi^{-2} \phi^{IJ}$, so that
$\mathcal{N}_I{}^J = 2\,F_{IK}(\hat\phi) \, \hat\phi^{KJ}$. With
respect to these, it is straightforward to evaluate \eqref{eq:FHG} for
$n=2$. In particular one can prove that $\mathcal{H}_{IJ} \equiv
\mathcal{H}_{IJ}^{(1)} = \tfrac{1}{8} \mathcal{H}_{IJ}^{(2)}$.  To
write down explicit Lagrangians, one may use an explicit
coordinatization of the $\phi^{IJ}$. This can, for instance, be done
by using the local $\mathrm{SU}(2)$ invariance to bring the matrix
$L^{rI}$ into lower triangular form,
\begin{equation} 
  \label{eq:explicit-L}
  L^{rI} = \frac{1}{\sqrt{2}} \,\left[ \begin{array}{ccc}
      0 & 0 & w_3 \\
      0 & u_2 & u_3 \\
      x_1 & x_2 & x_3
\end{array}
\right] \, ,
\end{equation}
so that 
\begin{equation}
  \label{eq:phi-wux}
  \begin{array}{lll}
      \phi^{11} = x_1{}^2 \,,\qquad & \phi^{22} = u_2{}^2 + x_2{}^2 
      \,,\qquad & \phi^{33} = w_3{}^2 + u_3{}^2 + x_3{}^2 \,, \\ 
     \phi^{12} = x_1\,x_2 \,,\qquad &\phi^{13} = x_1\,x_3 \,,
     \qquad & \phi^{23}  = u_2 \,u_3 + x_2\,x_3 \, . 
\end{array}
\end{equation}
Note that the values of the scalars $\phi^{IJ}$ are constrained to the
hypersurface $\tr(\cN) = 1$. This constraint can locally be solved for
one of the coordinates defined in (\ref{eq:phi-wux}) which is then
expressed in terms of the other scalars.

Using these results, the evaluation of the general tensor multiplet
Lagrangian \eqref{eq:tensor-boson-noV} yields
\begin{eqnarray}
  \label{eq:L-n2}
  e^{-1} \mathcal{L}  &=& 
  \tp \Big[ \tfrac{1}{6} R  
     +  \tfrac1{2} D  -\tfrac1{4} (\partial_\mu
    \ln \tp)^2 \Big] \nn\\[1ex]    
    &&{} 
    - \tp \,
    \Big[\mathcal{G}^{(1)}_{IJ} \,\partial_\mu\vec L^I \cdot 
    \partial^\mu\vec L^{J} + 
    \mathcal{G}^{(2)}_{IJ,KL} \, (\vec L^I\cdot\partial_\mu \vec L^J) \, 
    (\vec L^K\cdot\partial^\mu \vec L^L) \Big] 
    \nonumber\\[1ex]
    &&{}
    + \tp^{-1}\, \mathcal{H}_{IJ}\,
    E_{\mu}{}^I  E^{\mu J} 
    + E^{\mu I} \, \left[8 \, \mathcal{H}_{IJ}\, \vec L^J\cdot(\vec
    L^K\times\partial_\mu\vec L^L) \, F_{KL} + A_{\mu I} \right] \, .
\end{eqnarray}

In the final step we introduce Lagrange multipliers $s_I$ to impose
the Bianchi identity on the field strength $E^{\mu I}$ and eliminate
these field strengths as unconstrained fields through their equations
of motions. This results in the following eight-dimensional
quaternion-K\"ahler metric, 
\begin{equation} 
  \label{eq:le}
\begin{split}
{\rm d}s^2 = & \, \mathcal{K}^{(1)}_{IJ} \, {\rm d} \vec L^I \cdot
    {\rm d} \vec L^{J} + 
    \mathcal{K}^{(2)}_{IJ,KL} \, (\vec L^I\cdot {\rm d} \vec L^J) \, 
    (\vec L^K\cdot {\rm d} \vec L^L) \\ & \,
        +4 \, (A_{I} + {\rm d} s_I) \, \vec{L}^I \cdot (\vec{L}^K
    \times {\rm d} \vec{L}^L) F_{KL}  
    +\tfrac{1}{4}\big[\mathcal{H}^{-1} \big]^{IJ} (A_{I} + {\rm d}
    s_I) \, (A_{J} + {\rm d} s_J) \, ,  
\end{split}
\end{equation}
with the matrices
\begin{eqnarray} 
\mathcal{K}^{(1)}_{IJ} & = & F_{IJ} - \frac{1}{384 \, \det(\mathbf{M})}
\Big[ 1 + 3 \, \tr(\mathcal{N}^2) - 4 \, \tr(\mathcal{N}^3)
\Big] \, \mathcal{N}_{IJ} \, , \nn \\ 
\mathcal{K}^{(2)}_{IJ,KL} &= & 
4 \, F_{IL} \, F_{JK} - \frac{1}{64  \, \det(\mathbf{M})} \Big[
(1 - \tr(\cN^2) ) \, (F_{IK} \, \cN_{JL} + F_{JL} \, \cN_{IK})
\nn \\ && 
\qquad \qquad \quad
+ 2 \, ( \cN^2_{IK} \, \cN_{JL} + \cN^2_{JL} \, \cN_{IK} )
-4 \, \cN_{IK} \, \cN_{JL} 
 \Big] \, .
\end{eqnarray}
Here we suppressed the overall factor of $\chi$ which, when frozen to
a constant value, sets the curvature scale of the metric. In view of
the constraint, the $\phi^{IJ}$ depend on 5 coordinates. The extra 3
coordinates are provided by the fields $s^I$ originating from the
tensor sector of the model.  The line-element \eqref{eq:le} thus 
provides the extension of the classification of four-dimensional toric 
quaternion-K\"ahler manifolds by Calderbank and Pedersen \cite{CP} to
eight dimensions.

\section*{Acknowledgement}
  We are grateful to Stefan Vandoren for valuable discussions.  We thank
  the organizers of the RTN ForcesUniverse Network Workshop for a
  stimulating meeting. F.S.\ is supported by a European Commission
  Marie Curie Postdoctoral Fellowship under contract number
  MEIF-CT-2005-023966.  This work is partly supported by EU contract
  MRTN-CT-2004-005104, by INTAS contract 03-51-6346, and by NWO grant
  047017015.

\end{document}